\documentclass[%
 reprint,
amsmath,amssymb,
 aps,
]{revtex4-2}

\usepackage{graphicx}
\usepackage{dcolumn}
\usepackage{bm}
\usepackage{physics}
\usepackage{braket}
\usepackage{amsmath,amssymb,amstext,mathtools}

\usepackage[pdftex,pagebackref=false]{hyperref} 
\hypersetup{
    plainpages=false,       
    unicode=false,          
    pdftoolbar=true,        
    pdfmenubar=true,        
    pdffitwindow=false,     
    pdfstartview={FitH},    
    pdfnewwindow=true,      
    colorlinks=true,        
    linkcolor=blue,         
    citecolor=green,        
    filecolor=magenta,      
    urlcolor=cyan           
}

\begin{document}

\preprint{APS/123-QED}
\title{Achieving $10^{-5}$ level relative intensity crosstalk in optical holographic qubit addressing via a double-pass digital micromirror device}
\author{Shilpa Mahato}
\author{Rajibul Islam}
\affiliation{%
 Institute for Quantum Computing and Department of Physics and Astronomy, University of Waterloo, Waterloo, ON, N2L 3G1, Canada
}%












\begin{abstract}
Holographic beam shaping is a powerful approach for generating individually addressable optical spots for controlling atomic qubits, such as those in trapped-ion quantum processors.
However, its application in qubit control is limited by residual intensity crosstalk at neighboring sites and by a nonzero background floor in the far wings of the addressing beam, leading to accumulated errors from many exposed qubits.
Here, we present an all-optical scheme that mitigates both effects using a single digital micromirror device (DMD) operated in a double-pass configuration, in which light interacts with two separate regions of the same device.
In the first pass, one region of the DMD is placed in a Fourier plane and implements a binary-amplitude hologram for individual addressing, while in the second pass a different region serves as a programmable intermediate image-plane aperture for spatial filtering.
By multiplexing the Fourier-plane hologram to include secondary holograms, we generate weak auxiliary fields that interfere destructively with unwanted light at selected sites, while image-plane filtering suppresses the residual tail at larger distances.
Together, these techniques maintain relative intensity crosstalk at or below $10^{-5}$ ($-50\,\mathrm{dB}$) across the full field of view relevant for qubit addressing, and further reduce the far-wing background to approximately $10^{-6}$ at large distances from the addressed qubit, approaching the detection limit.
These results provide a compact, DMD-based solution for low-crosstalk optical holographic qubit addressing that is directly applicable to trapped ions and other spatially ordered quantum systems.
\end{abstract}

\maketitle

\section{Introduction}
Holographic beam shaping is a widely used technique for generating high-precision, individually addressable beams with applications in optogenetics \cite{Pegard2017}, lithography \cite{Ouyang2023, Jenness2008, Kelemen2007}, three-dimensional printing \cite{Lee2021}, micro-machining \cite{Zhang2016, Kuang2009}, and optical trapping \cite{Grier:06}.
In this approach, a programmable amplitude or phase hologram implemented through adaptive optics in the Fourier plane is used to engineer a target wavefront \cite{Zupancic2016UltraPrecise, Cizmar2010InsituWavefront}, which is then projected through a focusing lens or microscope objective to create spatially resolved addressing patterns.
This versatility also extends to quantum science, where holographic methods have become indispensable for controlling atomic systems with high spatial precision \cite{Motlakunta2024, Zhang:24}.

Holographic individual addressing has been demonstrated in trapped ions and in neutral atoms in both electronic ground and Rydberg states, enabling a variety of quantum-technology applications \cite{Motlakunta2024, Kim2016, Labuhn2016Tunable2D, Bernien2017ProbingManyBody}.
The precision afforded by this approach helps minimize unwanted intensity crosstalk, which is essential for targeting a single qubit without perturbing its neighbors.
Aberrations in the optical path can nevertheless degrade spot quality, and in situ characterization—either using external sensors or the atomic system itself—has been essential for compensating such distortions.
Both LCOS spatial light modulators and digital micromirror devices (DMDs) have been used to implement holographic control; the latter offer significant speed advantages, making them attractive for trapped-ion processors despite their binary-amplitude constraint. 
With aberration compensation \cite{Zupancic2016UltraPrecise} and hologram-optimization algorithms such as the iterative Fourier transform algorithm (IFTA) \cite{Wyrowski1989IterativeQuantization} the relative intensity crosstalk, $I_X$ at the \(10^{-4}\) or $- 40$ dB level at the neighboring qubit location has been achieved in trapped-ion systems \cite{Shih2021, Motlakunta2024, Chen:26}, demonstrating the feasibility of demanding operations such as in situ mid-circuit measurement and resets of quantum information.

\begin{figure*}
    \includegraphics[width=0.9\linewidth]{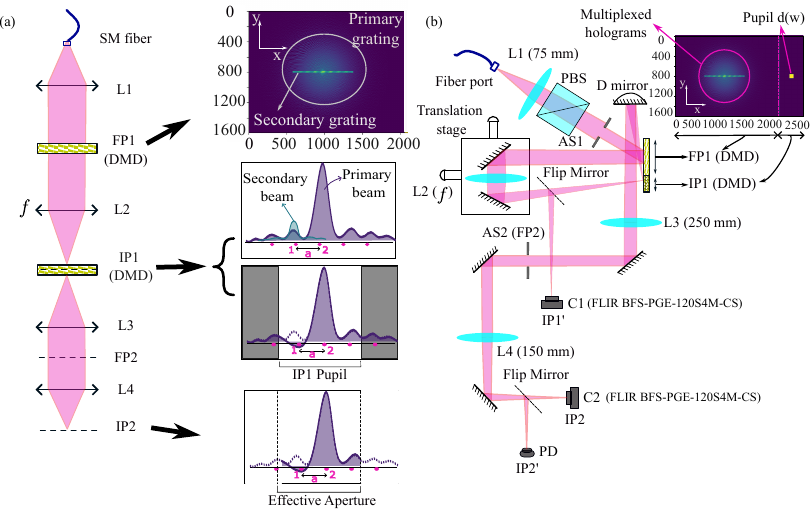}
    \caption{\textbf{Schematic of the experimental setup.}
    \textbf{(a)} A simplified optical layout. \textit{Left column:} Optical layout starting from a single-mode fiber delivering $\lambda = 370\,\mathrm{nm}$ light.
    Convex lenses are indicated by double arrows.
    Lens L1 collimates the beam to a Gaussian waist of $10\,\mathrm{mm}$ to illuminate a digital micromirror device (DMD), which acts as a Fourier plane (FP1) with respect to lens L2 (focal length $f$).
    The light interacts with the same DMD again at the image plane of L2 (IP1), and the field is relayed to the final image plane IP2.
    \textit{Right column:} \textit{(top)} Binary-amplitude holograms displayed on FP1 by toggling DMD micromirrors between ON (yellow) and OFF (purple) states (see text).
    \textit{(middle)} Schematic illustrating destructive interference between the fields generated by the primary and secondary holograms, suppressing relative intensity crosstalk at field point 1 while field point 2 is addressed.
    Distances between field points are expressed in units of the Gaussian beam waist at the corresponding image plane.
    A schematic of the DMD operation at IP1 shows its role as a programmable pupil for spatially filtering the tail of the illumination profile.
    \textit{(bottom)} Expected intensity profile at IP2, showing reduced crosstalk at field point 1 from destructive interference and additional suppression beyond an effective aperture set by image-plane filtering.
   \textbf{(b)} Detailed optical schematic of the DMD double-pass configuration.
   Focal length of lenses L1, L2, L3, L4 are shown in paranthesis.
    A translation stage is used to finely adjust the position of lens L2 so that the DMD is simultaneously located at both the Fourier plane and the image plane of L2.
    AS1 and AS2 are irises acting as Fourier-plane apertures, with AS1 ensuring selective illumination of FP1 on the DMD and AS2 blocking unwanted diffraction orders generated during the second pass.
    Planes $\mathrm{IP1}'$ and $\mathrm{IP2}'$ are equivalent to IP1 and IP2, respectively, and are used for monitoring the optical field.
    C1 and C2 denote CMOS cameras, and PD denotes a photodiode.
    \textit{(Top right)} The DMD is shown spatially partitioned into FP1 and IP1 regions, with the pink dashed line marking the spatial demarcation between FP1 and IP1. 
    }
    \label{fig:Schematic}
\end{figure*}

Despite the substantial progress enabled by aberration compensation and advanced hologram-generation algorithms, two limitations of existing holographic approaches are major roadblocks for scaling up their applications in quantum information processing.
First, further suppressing nearest-neighbor crosstalk significantly below the $-40$ dB level has proven challenging in practice and becomes an important requirement for more precise qubit operations and scaling up these processors. 
Second, the intensity profile typically exhibits a finite residual background leading to a finite crosstalk floor at large distances from the addressed site, rather than decaying to zero.
This background illumination can accumulate across many qubits, contributing to additional error that increases with increasing system size.




In this work, we address both of these limitations in the context of a trapped-ion quantum processor, where individual ytterbium ions are addressed with a $\lambda=370\,\mathrm{nm}$ laser beam.
Programmable addressing patterns are generated using a binary-amplitude hologram computed with IFTA and displayed on a DMD.
This hologram implements site-resolved control along the ion chain and serves as the foundation on which we introduce two complementary techniques for enhanced crosstalk suppression.

To mitigate crosstalk at sites proximal to the target qubit, we partition the Fourier plane into two regions that together form a composite hologram.
Most of the plane retains the original IFTA-optimized pattern, while a small subsection is replaced with a secondary hologram, in the form of a binary grating, engineered to produce a weak auxiliary optical field at a chosen neighboring site.
By adjusting the phase and amplitude of this auxiliary field, we induce destructive interference with the primary beam, i.e., the optical field generated by the IFTA hologram, creating a local intensity minimum.
This controlled interference yields a relative suppression of up to \(10\ \mathrm{dB}\), reducing $I_X$ to approximately $-50$ dB relative to the peak intensity, at a distance of 4$w$ from the target addressing beam of Gaussian waist $w$.
The Fourier hologram can be further multiplexed to add more such secondary holograms, with each hologram optimized to lower intensity crosstalk at a specific off-target qubit location.
This approach is conceptually along the line of Ref. \cite{Flannery2024PhysicalCoherentCancellation} but in a different setting.

To reduce the finite crosstalk background at far distances, we introduce an intermediate image plane in the optical path, where we apply a programmable field aperture that removes the weak residual tail of the addressing beam while preserving the primary spot.
This functionality is enabled by reusing the same DMD in a double-pass configuration: in the first pass, one region of the device acts as the multiplexed Fourier-plane hologram, while in the second pass a separate region is placed in an intermediate image plane and implements the spatial filter.
In this work, the term ``double-pass" refers to this two-step interaction of the beam with different regions of the same reflective DMD.
We find that $I_X$ falls below \(-50\,\mathrm{dB}\) at distances larger than about \(14w\) from the target, and approaches $-60$ dB beyond $30w$.
Our strategy of combining multiplexed Fourier holograms with intermediate image-plane filtering on the same DMD enables substantially reduced crosstalk at both near and far distances from the target qubit, providing a pathway toward higher-fidelity and scalable optical control in trapped-ion and other atomic quantum processors. 
\begin{figure*}
    \centering
    \includegraphics[width=0.9\linewidth]{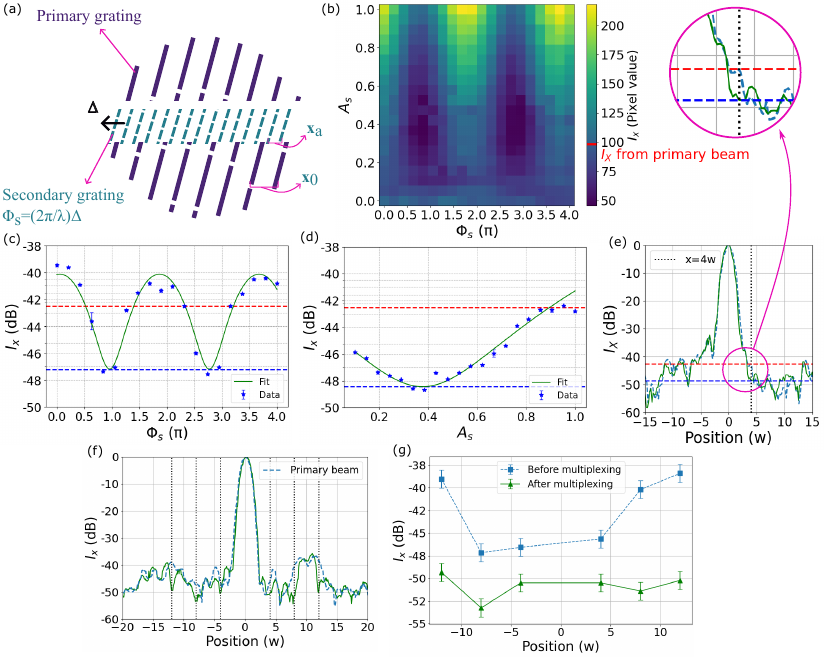}
            \caption{\textbf{Multiplexing Fourier holograms.} 
         \textbf{(a)} Illustration of multiplexing a secondary grating hologram with a primary grating hologram.
    $\mathbf{x}_0$ and $\mathbf{x}_a$ denote the spatial-frequency vectors of the primary and secondary gratings, respectively.
    The grating amplitude is controlled by the density of DMD mirrors in the ON state, while the phase is adjusted by lateral displacement of the grating.
    Unless otherwise noted, data in (b--e) were taken with $f = 150\,\mathrm{mm}$ and an IP1 beam waist of $w = 9\,\mathrm{\mu m}$ (see Fig.~\ref{fig:Schematic}); the PBS was not included.
    \textbf{(b)} Two-dimensional density plot of camera pixel values [0,255] measured at $4w$ from the target beam produced by the primary grating, shown as a function of secondary-hologram phase $\Phi_s$ and amplitude $A_s$.
    Data are averaged over three images acquired with a $1.2\,\mathrm{s}$ exposure on camera C1.
    \textbf{(c)} Relative intensity crosstalk $I_X$ (dB) versus $\Phi_s$ at $4w$ for $A_s = 0.70$.
    The red dashed line indicates the primary-beam crosstalk ($I_X = -42.6\,\mathrm{dB}$).
    The blue dashed line shows the minimum extracted by fitting the linear-scale data to a cosine model.
    Error bars here and below denote the standard error estimated from the HDR images for each shot, with three shots acquired per measurement.
    \textbf{(d)} $I_X$ versus $A_s$ at $\Phi_s = 2.77\pi$, the optimized phase obtained from (c).
    Linear-scale data are fitted to a quadratic model and displayed on a logarithmic scale.
    Points with $A_s \ge 0.9$ are excluded from the fit due to significant deviation from the quadratic trend.
    The minimum of the fit yields the optimal value of $A_s = 0.38$.
    \textbf{(e)} Comparison of HDR images of the primary beam (blue dashed curve) and the beam obtained by multiplexing a secondary hologram with optimized $\Phi_s$ and $A_s$ (green curve), showing the creation of an intensity minimum at $4w$.
    Panels (f,g) show results for multiple secondary holograms, taken with $f = 250\,\mathrm{mm}$ and $w = 20\,\mathrm{\mu m}$.
    \textbf{(f)} HDR images averaged over five acquisitions.
    The blue dashed curve corresponds to the primary beam, while the green solid curve shows the beam generated by multiplexing secondary gratings to create intensity minima at $\pm 4w$, $\pm 8w$, and $\pm 12w$.
    \textbf{(g)} Relative intensity crosstalk, $I_X$ at different field points before and after multiplexing the secondary gratings shown in (f).
    }
    \label{fig:sgm}
\end{figure*}
\section{Results}
\subsection{Experimental Setup}
Our experimental setup is shown in Fig.~\ref{fig:Schematic}.
The $370\,\mathrm{nm}$ light emerging from a single-mode fiber is collimated and used to illuminate a DMD (Texas Instruments DLP9000XUV), which is placed in a Fourier plane with respect to a focusing lens L2 of focal length $f$.
The DMD has a micromirror pitch of $7.6\,\mathrm{\mu m}$, and each micromirror can be toggled between two states (ON/OFF) or corresponding tilt angles ($\pm 12^{\circ}$).
The active area of the DMD comprises $1600 \times 2560$ pixels, corresponding to a physical size of $12.1\,\mathrm{mm} \times 19.5\,\mathrm{mm}$.

The DMD interacts with the light twice, once in a Fourier plane and once in an image plane.
Figure~\ref{fig:Schematic}a presents a conceptual optical schematic, in which the two interactions with the DMD are shown linearly as separate elements—first as the Fourier-plane region (FP1) and then as the image-plane region (IP1)—along with the expected beam profiles at key planes in the optical path.
Figure~\ref{fig:Schematic}b shows the full schematic of the actual beam path implemented in the experiment.
In practice, the DMD is divided into two regions: a $1600 \times 2060$-pixel region (FP1), which acts as the Fourier-plane region, and the remaining $1600 \times 500$-pixel region (IP1), which acts as the image-plane region.
A flip mirror directs the light to a camera at $\mathrm{IP1}'$, which is used for aberration characterization and for verifying the operation of multiplexed Fourier holograms.
A two-lens imaging system with focal lengths $250\,\mathrm{mm}$ and $150\,\mathrm{mm}$ images the IP1 region of the DMD onto IP2 with a demagnification of $0.6$.
A D-shaped mirror is positioned to direct the diffracted orders after the second interaction with the DMD into this two-lens imaging system.
The aberration map is extracted at $\mathrm{IP1}'$ and IP2 using the method described in Ref.~\cite{Shih2021}, and a binary amplitude grating hologram generated with IFTA is displayed on the FP1 region of the DMD.
A second flip mirror directs the light to a photodiode at $\mathrm{IP2}'$, which is used to locate the beam center on the IP1 region of the DMD.
In practice, the image plane IP2 can be relayed onto an atomic system for quantum operations, although in this work we restrict our analysis to the optical system up to IP2.

\begin{figure*}
    \centering
    \includegraphics[width=1\linewidth]{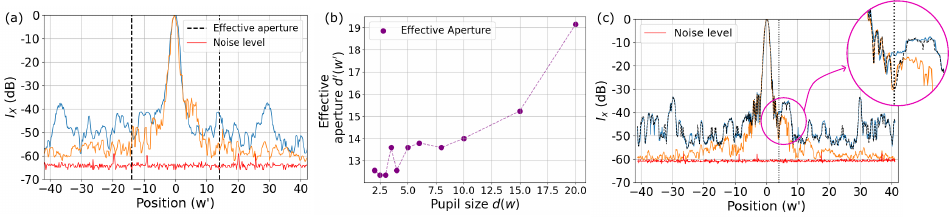}
    \caption{\textbf{Intermediate image plane filtering with DMD in double pass}. 
    The data in (a-c) were taken with $f$=250 mm. The red curve shows the background intensity floor when no light is being sent to the DMD.
    (\textbf{a}) HDR images (averaged over three acquisitions) of the primary beam in a singles pass (blue curve) at IP1' with waist, $w=9\ \mathrm{\mu m}$ and the beam after double pass (orange curve) with pupil size, $d=5w$ and waist, $w=15\ \mathrm{\mu m}\implies w'=9\ \mathrm{\mu m}$. The reflective pupil in the second pass results in an effective aperture corresponding to $\approx14w$. 
    (\textbf{b}) Each data point corresponds to the effective aperture ($d'$) extracted from the HDR images (averaged over three acquisitions) for different reflective pupil sizes ($d$). The effective aperture ($d'$) appears to saturate at $14w'$.  
    (\textbf{c})  The data were taken with $w=30\ \mathrm{\mu m}$, which results in $w'=18\ \mathrm{\mu m}$ and comprises HDR images averaged over five acquisitions. The blue curve shows the primary beam after the double pass with all mirrors of the IP1 (DMD) turned ON. The black dashed curve shows a beam generated from multiplexing a secondary grating with the primary grating at FP1 (DMD) that creates an intensity minimum at $4w$, reducing $I_X$ from $-40$ dB to $-50$ dB, followed by the double pass with all IP1 (DMD) mirrors turned ON. The green curve shows a beam generated by multiplexing a secondary hologram onto the primary hologram at FP1 (DMD), along with a reflective pupil ($d=6w$) turned on at the IP1 (DMD), resulting in an effective aperture of $8w$.}

    
    \label{fig:dp}
\end{figure*}
\subsection{Multiplexing the Fourier Plane}
In this section, we characterize the effect of Fourier-plane multiplexing using intensity measurements recorded at the first image plane, $\mathrm{IP1'}$, prior to image-plane filtering.
To suppress intensity crosstalk at selected locations, we multiplex the Fourier-plane hologram by introducing a localized secondary binary grating that generates a weak auxiliary field at a chosen image-plane position (Fig.~\ref{fig:sgm}a).
The secondary hologram is defined as
\begin{equation}
\label{eqn:secondary_grating}
\begin{split}
    F_\mathrm{s}(\mathbf{x})
    &= \eta\left|\frac{\mathrm{rect}\!\left(\frac{x}{l},\frac{y}{m}\right)}{E_\mathrm{in}(\mathbf{x})}\right|
    \frac{A_s}{2}
    \bigg(
        \cos\Big(
            2\pi \frac{\mathbf{x_a}\cdot \mathbf{x}}{\lambda f}
            + \Phi_s
        \Big. \\
    &\hspace{5em}
        \Big.
            - \Phi_\mathrm{in}(\mathbf{x})
        \Big)
        + 1
    \bigg).
\end{split}
\end{equation}

Here, $\lvert E_\mathrm{in}(\mathbf{x})\rvert$ is the amplitude of the input field at location $\mathbf{x}\equiv(x,y)$ in the DMD Fourier plane, and $\mathrm{rect}\!\left(\frac{x}{l},\frac{y}{m}\right)$ defines the rectangular window over which the secondary grating is applied.
The vector $\mathbf{x_a}$ is the spatial-frequency vector chosen such that the secondary grating produces an auxiliary beam centered at $\mathbf{X}_0+\mathbf{a}$ in the image plane, where $\mathbf{X}_0$ denotes the target-beam location and $\mathbf{a}$ specifies the displacement at which crosstalk suppression is desired.
The term $\Phi_\mathrm{in}(\mathbf{x})$ is the input phase map obtained from in situ aberration characterization (Ref.~\cite{Shih2021}), and the factor $1/\lvert E_\mathrm{in}(\mathbf{x})\rvert$ compensates for spatial nonuniformity of the incident field across the DMD.
The normalization factor $\eta$ enforces $0\leq F_\mathrm{s}(\mathbf{x})\leq 1$.
Finally, $A_s$ and $\Phi_s$ are the amplitude and phase of the secondary hologram, with $0\leq A_s\leq 1$.

After measuring the baseline crosstalk $I_X$ at the desired location from the primary beam alone, we choose the window size $(m\times l)$ to provide sufficient dynamic range to tune the auxiliary-beam amplitude through $A_s$ while minimizing disturbance to the primary IFTA hologram.
For trapped-ion addressing in a linear chain, we choose $m=2$ DMD pixels and $l=460$ DMD pixels.
This choice reflects the fact that suppression is required primarily along the ion-chain axis, while the large transverse spread from the narrow width in the Fourier plane is acceptable provided it does not illuminate nearby structures such as trap electrodes.

The secondary grating is generated using a random binarization method \cite{Zupancic2016UltraPrecise}.
To identify the optimum interference condition, we keep the primary IFTA hologram fixed and scan the secondary-hologram phase $\Phi_s$ while the secondary hologram is applied.
Figure~\ref{fig:sgm}b shows camera images recorded at $\mathrm{IP1'}$ as both $\Phi_s$ and $A_s$ are varied in a grid search, revealing regions of reduced intensity due to destructive interference between the primary field and the auxiliary field.
For a fixed $A_s$, we fit the phase scan (Fig.~\ref{fig:sgm}c) to determine the optimum $\Phi_s$, and then maximize the suppression by scanning $A_s$ at that phase (Fig.~\ref{fig:sgm}d).
Using this procedure, the secondary hologram reduces the crosstalk to $-48^{+1}_{-2}\,\mathrm{dB}$, an average improvement of about $5\,\mathrm{dB}$ over the baseline crosstalk of $-42.6\,\mathrm{dB}$ obtained with the primary hologram alone (Fig.~\ref{fig:sgm}e). 
The uncertainty in the reduced crosstalk value was determined from the fitting errors of the parameters obtained from the amplitude fit curve shown in Fig. \ref{fig:sgm}d.

We next extend this approach by employing multiple secondary holograms to suppress $I_X$ at several locations simultaneously.
In this demonstration, we introduce six secondary holograms in the Fourier-plane region to minimize $I_X$ at displacements of $4w$, $8w$, and $12w$ on either side of the addressed site.
We first estimate the required auxiliary-field amplitudes from the measured $I_X$ values of the primary beam at the corresponding locations.
When two secondary holograms require comparable amplitudes, we overlay them on the same subset of pixels to remove as little of the primary IFTA hologram as possible.
However, we avoid overlaying more than two secondary holograms on the same DMD region, as doing so degrades independent control of their relative amplitudes and phases.
When more than two secondary holograms are required, or when their amplitudes differ significantly, we spatially separate the corresponding holograms by stacking their windows along the $y$ direction in the Fourier plane.
We then scan the phase of each secondary hologram independently, followed by a fine amplitude scan of each hologram at its optimum phase.
Using this multiplexed strategy, we achieve $I_X \lesssim -50\,\mathrm{dB}$ at all six target sites (Fig.~\ref{fig:sgm}f-g).
The small difference between the crosstalk obtained with a single secondary hologram and that achieved using multiplexed secondary holograms arises from differences in experimental conditions between the two measurements, including the beam waist and polarization optics, as detailed in the caption Fig.~\ref{fig:sgm}.

\subsection{DMD in Double Pass}
In this subsection, we characterize image-plane filtering using intensity measurements recorded at IP2 with a camera.
This method uses the optical setup shown in Fig.~\ref{fig:Schematic} up to the plane IP2.
The addressing beam is generated from a hologram displayed on the Fourier-plane region of the DMD, computed with IFTA and incorporating an aberration map characterized up to IP2.
In our configuration, the Fourier-plane hologram occupies a $1600 \times 2060$-pixel region (FP1) of the DMD, while the remaining $1600 \times 500$-pixel region (IP1) is used as an intermediate image plane.

The location of the beam center at IP1 on the DMD is calibrated in two steps: a coarse search that narrows down the beam location using successively smaller blocks of mirrors, followed by a fine linear scan in which individual micromirrors are turned on one by one and the center is determined from the camera signal at IP2.
By turning on different numbers of mirrors centered on this beam spot, we realize a programmable reflective square pupil of size $d$ that acts as a field aperture and filters the unwanted tail arising from the primary beam spot.
Aberrations and diffraction in the relay optics between IP1 and IP2 broaden the imaged pupil and soften its edge relative to the geometric-optics prediction based on the relay magnification.

To quantify the filtering performance, we define an effective aperture size $d'$ as the distance along the ion axis at which the relative intensity crosstalk $I_X$ falls below $-50\,\mathrm{dB}$.
This definition is chosen as a convenient benchmark, and other thresholds can be used equivalently.
We first investigate image-plane filtering for a beam created by IFTA without the addition of any secondary hologram (Fig.~\ref{fig:dp}a).
We observe that the effective aperture size $d'$ decreases as the pupil size $d$ on IP1 is reduced, but $d'$ saturates at approximately $14w'$, where $w'$ is the beam waist at IP2 (Fig.~\ref{fig:dp}b).

This observation motivates a combined approach in which Fourier-plane multiplexing suppresses crosstalk at near locations (within $d'$), while image-plane filtering mitigates the residual tail at larger distances.
As an example, Fig.~\ref{fig:dp}c shows the combination of a secondary grating targeting the location at $4w'$ with image-plane filtering, demonstrating interferometric cancellation at a neighboring site together with suppression at faraway locations.
With the combined approach, the relative intensity crosstalk is maintained at or below $-50\,\mathrm{dB}$ across the full field, and it decreases by an additional order of magnitude to near $-60\,\mathrm{dB}$ beyond approximately $30w'$, approaching the detector noise floor within measurement fluctuations.





\section{Conclusion}

In summary, we have demonstrated relative intensity crosstalk below the $10^{-5}$ level ($-50\,\mathrm{dB}$) for optical addressing of trapped-ion qubits using a double-pass digital micromirror device.
This performance is achieved through a dual strategy that combines Fourier-plane multiplexing with additional holograms optimized to interferometrically cancel unwanted optical fields at selected locations, together with programmable filtering in an intermediate image plane.
At large distances from the addressed qubit, image-plane filtering further suppresses residual background light, reducing the crosstalk to the $10^{-6}$ level.
These results enable more precise qubit manipulation, including coherent control as well as mid-circuit measurement and reset operations, which are essential for scaling up quantum information processing.

The approach presented here can be readily extended to address multiple qubits simultaneously within the same optical architecture and without hardware modification.
Multiple pupils can be implemented in the intermediate image plane to independently filter beams addressing well-separated qubits, while a single pupil can be used to cover multiple nearby qubits when combined with multiplexed secondary holograms.

In this manuscript, we have presented results on Fourier-plane multiplexing with up to six secondary holograms.
We find that as the number of secondary holograms increases, overlap between the auxiliary beams generated by each hologram becomes significant, leading to increased crosstalk at the target sites.
This overlap arises in part from our choice of a rectangular window in Eq.~\ref{eqn:secondary_grating}, which introduces diffraction sidelobes away from the centers of the auxiliary beams.
This effect could be mitigated by smoothing the secondary-grating window.
In addition, increasing the number of secondary holograms degrades the primary addressing beam by removing an increasing fraction of its Fourier components.
Experimentally, we observe that the crosstalk performance at the target sites degrades noticeably when the number of secondary holograms exceeds approximately eight, although a more careful analysis is required to quantitatively determine the limits on the number of secondary holograms.

We also find that the quality of the beam in the image plane degrades with the introduction of the intermediate image plane.
Specifically, the relative intensity crosstalk increases by approximately $6\,\mathrm{dB}$ at four times the respective beam waists at IP1 and IP2 when comparing Figs.~\ref{fig:sgm}f and \ref{fig:dp}a.
This behavior is likely due to the small beam waist at IP1, which is on the order of two DMD pixels, making it difficult to block multiple diffraction orders from the DMD mirrors at IP1 while allowing the desired Fourier components to pass through the aperture AS2 in FP2 (Fig.~\ref{fig:Schematic}b).
In addition, the crosstalk is affected by scattering from the DMD mirror surfaces in IP1.
Furthermore, since the mirrors on the DMD do not form a continuous plane when turned on, they introduce additional aberrations \cite{Zhang:24}.
Despite the increased near-field noise introduced by the intermediate image plane, the secondary holograms are able to cancel the resulting crosstalk through destructive interference, highlighting the robustness of our approach.

The performance of the intermediate image-plane filtering could be further improved by introducing aberration compensation between IP1 and IP2, for example using a deformable mirror.
Finally, the beam can be relayed onto ion qubits, and the ions themselves can be used for fine-tuning aberrations, optimizing secondary gratings, and performing high-dynamic-range intensity calibration \cite{Motlakunta2024}.
This hybrid approach enables optimization beyond what is possible with classical imaging alone and provides a practical pathway for deploying the techniques demonstrated here in operating trapped-ion quantum processors.

\section*{Acknowledgments}
We thank Chung-You Shih for suggesting the possibility of combining both the Fourier plane and the image plane on the same device.
We acknowledge financial support from the Canada First Research Excellence Fund (CFREF); the Natural Sciences and Engineering Research Council of Canada (NSERC); the University of Waterloo; and Innovation, Science and Economic Development Canada (ISED).

\section*{Author Contributions}
S.M.\ built the experimental setup and designed the measurement protocols based on initial ideas from R.I.
S.M.\ acquired and analyzed all data.
S.M.\ and R.I.\ contributed to scientific discussions and to the writing of the manuscript.
R.I.\ supervised the entire project.

\bibliography{references}

@article{Pegard2017,
  title   = {Three-dimensional scanless holographic optogenetics with temporal focusing (3D-SHOT)},
  author  = {P\'egard, Nicolas C. and Mardinly, Alan R. and Oldenburg, Ian Ant{\'o}n and Sridharan, Savitha and Waller, Laura and Adesnik, Hillel},
  journal = {Nature Communications},
  volume  = {8},
  number  = {1},
  pages   = {1228},
  year    = {2017},
  publisher = {Nature Publishing Group},
  doi     = {10.1038/s41467-017-01031-3}
}

@article{Ouyang2023,
  author    = {Ouyang, Wenqi and Xu, Xiayi and Lu, Wanping and Zhao, Ni and Han, Fei and Chen, Shih-Chi},
  title     = {Ultrafast 3D nanofabrication via digital holography},
  journal   = {Nature Communications},
  volume    = {14},
  number    = {1},
  pages     = {1716},
  year      = {2023},
  doi       = {10.1038/s41467-023-37163-y}
}

@article{Jenness2008,
  author    = {Jenness, N.~J. and Wulff, K.~D. and Johannes, M.~S. and Padgett, M.~J. and Cole, D.~G. and Clark, R.~L.},
  title     = {Three-dimensional parallel holographic micropatterning using a spatial light modulator},
  journal   = {Optics Express},
  volume    = {16},
  number    = {20},
  pages     = {15942--15948},
  year      = {2008},
  doi       = {10.1364/OE.16.015942}
}

@article{Kelemen2007,
  author    = {Kelemen, L{\'o}r{\'a}nd and Valkai, S{\'a}ndor and Ormos, P{\'a}l},
  title     = {Parallel photopolymerisation with complex light patterns generated by diffractive optical elements},
  journal   = {Optics Express},
  volume    = {15},
  number    = {22},
  pages     = {14488--14497},
  year      = {2007},
  doi       = {10.1364/OE.15.014488}
}

@article{Lee2021,
  author    = {Lee, Chung-Fei and Hsu, Wei-Feng and Yang, Tzu-Hsuan and Chung, Ren-Jei},
  title     = {{Three-Dimensional (3D) Printing Implemented by Computer-Generated Holograms for Generation of 3D Layered Images in Optical Near Field}},
  journal   = {Photonics},
  volume    = {8},
  number    = {7},
  pages     = {286},
  year      = {2021},
  doi       = {10.3390/photonics8070286}
}

@article{Zhang2016,
  author    = {Zhang, Chenchu and Hu, Yanlei and Du, Wenqiang and Wu, Peichao and Rao, Shenglong and Cai, Ze and Lao, Zhaoxin and Xu, Bing and Ni, Jincheng and Li, Jiawen and Zhao, Gang and Wu, Dong and Chu, Jiaru and Sugioka, Koji},
  title     = {Optimized holographic femtosecond laser patterning method towards rapid integration of high-quality functional devices in microchannels},
  journal   = {Scientific Reports},
  volume    = {6},
  number    = {1},
  pages     = {33281},
  year      = {2016},
  doi       = {10.1038/srep33281},
  url       = {https://doi.org/10.1038/srep33281}
}

@article{Kuang2009,
  author    = {Kuang, Zheng and Liu, Dun and Perrie, Walter and Edwardson, Stuart and Sharp, Martin and Fearon, Eamonn and Dearden, Geoff and Watkins, Ken},
  title     = {Fast parallel diffractive multi-beam femtosecond laser surface micro-structuring},
  journal   = {Applied Surface Science},
  volume    = {255},
  number    = {13-14},
  pages     = {6582--6588},
  year      = {2009},
  doi       = {10.1016/j.apsusc.2009.02.043}
}

@article{Grier:06,
author = {David G. Grier and Yael Roichman},
journal = {Appl. Opt.},
number = {5},
pages = {880--887},
publisher = {Optica Publishing Group},
title = {Holographic optical trapping},
volume = {45},
month = {Feb},
year = {2006},
url = {https://opg.optica.org/ao/abstract.cfm?URI=ao-45-5-880},
doi = {10.1364/AO.45.000880},
}

@article{Motlakunta2024,
  author = {Motlakunta, Sainath and Kotibhaskar, Nikhil and Shih, Chung-You and Vogliano, Anthony and McLaren, Darian and Hahn, Lewis and Zhu, Jingwen and Hablützel, Roland and Islam, Rajibul},
  title = {Preserving a qubit during state-destroying operations on an adjacent qubit at a few micrometers distance},
  journal = {Nature Communications},
  volume = {15},
  number = {1},
  pages = {6575},
  year = {2024},
  doi = {10.1038/s41467-024-50864-2},
  url = {https://doi.org/10.1038/s41467-024-50864-2},
}

@article{Shih2021,
  author = {Shih, Chung-You and Motlakunta, Sainath and Kotibhaskar, Nikhil and Sajjan, Manas and Hablützel, Roland and Islam, Rajibul},
  title = {Reprogrammable and high-precision holographic optical addressing of trapped ions for scalable quantum control},
  journal = {npj Quantum Information},
  volume = {7},
  number = {1},
  pages = {57},
  year = {2021},
  doi = {10.1038/s41534-021-00396-0},
  url = {https://doi.org/10.1038/s41534-021-00396-0}
}

@article{Zhang:24,
author = {Bichen Zhang and Pai Peng and Aditya Paul and Jeff D. Thompson},
journal = {Optica},
number = {2},
pages = {227--233},
publisher = {Optica Publishing Group},
title = {Scaled local gate controller for optically addressed qubits},
volume = {11},
month = {Feb},
year = {2024},
url = {https://opg.optica.org/optica/abstract.cfm?URI=optica-11-2-227},
doi = {10.1364/OPTICA.512155},
}

@article{Kim2016,
  author    = {Hyosub Kim and Woojun Lee and Han-gyeol Lee and Hanlae Jo and 
               Yunheung Song and Jaewook Ahn},
  title     = {In situ single-atom array synthesis using dynamic holographic optical tweezers},
  journal   = {Nature Communications},
  volume    = {7},
  number    = {1},
  pages     = {13317},
  year      = {2016},
  doi       = {10.1038/ncomms13317},
  url       = {https://doi.org/10.1038/ncomms13317},
}

@article{Zupancic2016UltraPrecise,
  author  = {Zupancic, P. and Yefsah, T. and Desbuquois, R. and Nascimb{\`e}ne, S. and Aidelsburger, M. and Dalibard, J.},
  title   = {Ultra-precise holographic beam shaping for microscopic quantum control},
  journal = {Optics Express},
  volume  = {24},
  number  = {13},
  pages   = {13881--13893},
  year    = {2016},
  doi     = {10.1364/OE.24.013881}
}

@article{Cizmar2010InsituWavefront,
  author  = {{\v{C}}i{\v{z}}m{\'a}r, T. and Mazilu, M. and Dholakia, K.},
  title   = {In situ wavefront correction and its application to micromanipulation},
  journal = {Nature Photonics},
  volume  = {4},
  pages   = {388--394},
  year    = {2010},
  doi     = {10.1038/nphoton.2010.85}
}

@article{Wyrowski1989IterativeQuantization,
  author  = {Wyrowski, F.},
  title   = {Iterative quantization of digital amplitude holograms},
  journal = {Applied Optics},
  volume  = {28},
  number  = {18},
  pages   = {3864--3870},
  year    = {1989},
  doi     = {10.1364/AO.28.003864}
}

@article{Labuhn2016Tunable2D,
  author  = {Labuhn, Henning and Barredo, Daniel and Ravets, Sylvain and de L{\'e}s{\'e}leuc, Sylvain and Macr{\`i}, Tommaso and Lahaye, Thierry and Browaeys, Antoine},
  title   = {Tunable two-dimensional arrays of single Rydberg atoms for realizing quantum Ising models},
  journal = {Nature},
  volume  = {534},
  number  = {7609},
  pages   = {667--670},
  year    = {2016},
  doi     = {10.1038/nature18274}
}

@article{Bernien2017ProbingManyBody,
  author  = {Bernien, Hannes and Schwartz, Sylvain and Keesling, Alexander and Levine, Harry and Omran, Ahmed and Pichler, Hannes and Choi, Soonwon and Zibrov, Alexander S. and Endres, Manuel and Greiner, Markus and Vuleti{\'c}, Vladan and Lukin, Mikhail D.},
  title   = {Probing many-body dynamics on a 51-atom quantum simulator},
  journal = {Nature},
  volume  = {551},
  number  = {7682},
  pages   = {579--584},
  year    = {2017},
  doi     = {10.1038/nature24622}
}

@article{Chen:26,
author = {Yan Chen and Quan Long and Yifan Zhou and Enteng An and Ran He and Jinming Cui and Yunfeng Huang and Chuanfeng Li},
journal = {Chin. Opt. Lett.},
number = {1},
pages = {010201},
publisher = {Optica Publishing Group},
title = {In-situ aberration correction for far-detuned laser systems via a trapped ion probe},
volume = {24},
month = {Jan},
year = {2026},
url = {https://opg.optica.org/col/abstract.cfm?URI=col-24-1-010201},
}

@article{Flannery2024PhysicalCoherentCancellation,
  title        = {Physical coherent cancellation of optical addressing crosstalk in a trapped-ion experiment},
  author       = {Flannery, Jeremy and Matt, Roland and Huber, Luca Immanuel and Wang, Kaizhao and Axline, Christopher James and Oswald, Robin and Home, Jonathan P.},
  journal      = {Quantum Science and Technology},
  volume       = {10},
  number       = {1},
  pages        = {015012},
  year         = {2025},
  doi          = {10.1088/2058-9565/ad8371},
  url          = {https://doi.org/10.1088/2058-9565/ad8371}
}

\end{document}